\documentclass[12pt]{article}
\usepackage{amsmath}
\usepackage{amsfonts}
\usepackage{amsthm}
\usepackage{amssymb}
\usepackage{bbold}
\usepackage{enumitem}

\DeclareSymbolFontAlphabet{\amsmathbb}{AMSb}

\newtheorem{theorem}{Fact}
\newcommand{\basis}[2]{{\boldsymbol\Theta}_{#2}^{#1}}

\title{Forces on an exterior algebra bundle}
\author{jason hanson}

\begin{document}
\maketitle

\begin{abstract}
In a previous article, the exterior algebra bundle over spacetime was used as a common geometric framework for obtaining the Dirac and Einstein equations, with other forces incorporated using minimal coupling.  Here the fundamental forces that are allowed within this framework are explicitly enumerated.
\end{abstract}

\section{Introduction}

In the article \cite{me}, we used the exterior algebra bundle $\bigwedge_\ast M$ over a spacetime manifold $M$ as a geometric framework for incorporating both the Dirac and Einstein equations.  We also indicated briefly how other forces can be introduced using minimal coupling, namely by using a connection on the exterior algebra bundle of the form $\nabla_\alpha=\partial_\alpha+\hat\Gamma_\alpha+\theta_\alpha$.  Here $\hat\Gamma_\alpha$ is the torsion--free metric--compatible connection on $M$ extended to $\bigwedge_\ast M$, and $\theta_\alpha$ is a collection of four $16\times 16$ matrices that encode the forces.  These matrices must satisfy certain constraints in order for the variational principle to yield the Dirac equation $\gamma^\alpha\nabla_\alpha\psi=m\psi$.  Such a collection ${\boldsymbol\theta}=\{\theta_0,\theta_1,\theta_2,\theta_3\}$ is a tensor, which we call a {\em force tensor} --- although it represents a potential rather than a force.

In this article, we describe the distinct types of forces that are allowed through minimal coupling.  This is done by considering the action of the Lorentz group on the space of all possible force tensors $V$.  Under this action, $V$ is a representation of the Lorentz group.  And as such, it can be decomposed into irreducible subrepresentations.  Each irreducible subrepresentation corresponds to a distinct force.

{\it Article summary.} In the remainder of this section, we review the relevant constructions used in the geometric framework introduced in \cite{me}, as well as the necessary constraints on a force tensor ${\boldsymbol\theta}$.  In section \ref{sec:space}, we use geodesic normal coordinates to find a basis for the space of all possible force tensors, and in section \ref{sec:irred} we write down the irreducible subspaces.  Each irreducible subspace is identified with a type of fundamental force field on spacetime $M$: scalar, vector, anti--symmetric tensor, symmetric tensor, and Fierz tensor.  In section \ref{sec:force}, we use the curvature of the connection $\nabla_\alpha$ to obtain field equations for each of the fundamental forces in a flat spacetime.

\subsection{Geometric framework}

Suppose $M$ is a Lorentz manifold with metric $g$.  Let $x=(x^\alpha)$, with $\alpha=0,1,2,3$, be local coordinates for $M$, and let ${\bf e}_\alpha\doteq\partial_\alpha=\partial/\partial x^\alpha$ denote the corresponding basis vectors of the tangent bundle of $M$ at $x$.  In this basis, $g_{\alpha\beta}$ denotes the components of $g$, and $g^{\alpha\beta}$ denotes the components of $g^{-1}$.  We assume that $g$ has signature $(+,-,-,-,-)$.

{\it Exterior algebra bundle.}
The exterior algebra bundle $\bigwedge_\ast M$ is formed by taking the complex exterior algebra of each fiber of the tangent bundle.  We may choose
$${\bf e}_\emptyset,\,
  {\bf e}_0,\,
  {\bf e}_1,\,
  {\bf e}_2,\,
  {\bf e}_3,\,
  {\bf e}_{01},\,
  {\bf e}_{02},\,
  {\bf e}_{03},\,
  {\bf e}_{12},\,
  {\bf e}_{13},\,
  {\bf e}_{23},\,
  {\bf e}_{012},\,
  {\bf e}_{013},\,
  {\bf e}_{023},\,
  {\bf e}_{123},\,
  {\bf e}_{0123}
$$
as basis vectors for the fibers of $\bigwedge_\ast M$.  Here ${\bf e}_I\doteq{\bf e}_{\alpha_1}\wedge{\bf e}_{\alpha_2}\wedge\cdots\wedge{\bf e}_{\alpha_k}$ for the multi--index $I=\alpha_1\alpha_2\cdots\alpha_k$.  We denote the length of $I$ by $|I|$; i.e., $|I|=k$.  A field $\psi$ is a section $M\rightarrow\bigwedge_\ast M$, and we write $\psi=\psi^I{\bf e}_I$, where $I=\emptyset,0,1,\dots,0123$.  We use $\bar\psi^I$ to denote the complex conjugate of $\psi^I$.

{\it Gamma matrices.}
In addition to the exterior product ${\bf e}_\alpha\wedge\psi$, we have the interior product $\iota_\alpha\psi$, which is the linear operator defined by the rules (i) $\iota_\alpha{\bf e}_\beta=g_{\alpha\beta}{\bf e}_\emptyset$ and (ii) $\iota_\alpha({\bf e}_I\wedge{\bf e}_J)=(\iota_\alpha{\bf e}_I)\wedge{\bf e}_J+(-1)^{|I|}{\bf e}_I\wedge(\iota_\alpha{\bf e}_J)$.  The gamma matrix $\gamma_\alpha$ is then defined to be the linear operator $\gamma_\alpha\psi\doteq{\bf e}_\alpha\psi-\iota_\alpha\psi$.  We find that $\gamma_\alpha\gamma_\beta+\gamma_\beta\gamma_\alpha=-2g_{\alpha\beta}$.  As usual, we set $\gamma^\alpha\doteq g^{\alpha\beta}\gamma_\beta$.

{\it Extended metric.}
The spacetime metric $g$ is extended to a Hermitian metric $\hat{g}$ on $\bigwedge_\ast M$ by setting $\hat{g}(\psi,\phi)=\bar\psi^I\hat{g}_{IJ}\psi^J$.  Here $\hat{g}_{IJ}=0$ if $|I|\neq|J|$, and $\hat{g}_{IJ}=\det(g_{\alpha_i\beta_j})$ if $I=\alpha_1\cdots\alpha_k$ and $J=\beta_1\cdots\beta_k$ with $k=|I|=|J|$.  One shows that $\gamma_\alpha^\dagger\hat{g}+\hat{g}\gamma_\alpha=0$.

{\it Extended group and algebra actions.}
A local transformation $A$ on the tangent bundle of $M$ extends to a transformation $\hat{A}$ on $\bigwedge_\ast M$ via the rule $\hat{A}{\bf e}_{\alpha_1\cdots\alpha_k}\doteq(A{\bf e}_{\alpha_1})\wedge\cdots\wedge(A{\bf e}_{\alpha_k})$.  In particular, a change of basis $B$ for the tangent bundle of $M$, ${\bf e}_\alpha'=B^{-1}{\bf e}_\alpha$, extends to a change of basis $\hat{B}$ on $\bigwedge_\ast M$.  If $G$ is a Lie group acting locally on the tangent bundle, then we can extend the action of $G$ to $\bigwedge_\ast M$ by using this rule.  On the other hand, the Lie algebra ${\mathfrak g}$ of $G$ extends to a Lie algebra action via the rule
\begin{equation}\label{eq:liealg}
  \hat{a}{\bf e}_{\alpha_1\cdots\alpha_k}
    =\sum_{i=1}^k{\bf e}_{\alpha_1}\wedge\cdots
       \wedge{\bf e}_{\alpha_{i-1}}\wedge(a{\bf e}_{\alpha_i})
       \wedge{\bf e}_{\alpha_{i+1}}\wedge\cdots
       \wedge{\bf e}_{\alpha_k}
\end{equation}
for all $a\in{\mathfrak g}$.  Note that $\hat{a}{\bf e}_\emptyset=0$.

\subsection{Allowed forces}

We use minimal coupling to model forces.  That is, we assume there is a connection $\nabla$ on $\bigwedge_\ast M$, where $\nabla_\alpha\psi=\partial_\alpha\psi+C_\alpha\psi$ for some collection of four $16\times 16$ matrices $C_\alpha$.  However, we need to ensure that we obtain the Dirac equation
$$D\psi=m\psi,
  \quad\text{where}\quad
  D\psi\doteq\gamma^\alpha\nabla_\alpha\psi,
$$
by varying the Lagrangian density
\begin{equation}\label{eq:diraclagrangian}
  L_K=(\tfrac{1}{2}\psi^\dagger\hat{g}D\psi+c.c.
       -m\psi^\dagger\hat{g}\psi)\,\omega,\quad
  \omega\doteq\sqrt{-\det(g)},
\end{equation}
with respect to $\bar\psi$.  As observed in \cite{me}, the conditions
\begin{equation}\label{eq:fullcompat}
  C_\alpha^\dagger\hat{g}+\hat{g}C_\alpha=\partial_\alpha\hat{g}
  \quad\text{and}\quad
  [\gamma^\alpha,C_\beta]
  =\partial_\beta\gamma^\alpha+\Gamma_{\beta\epsilon}^\alpha\gamma^\epsilon,
\end{equation}
where $\Gamma_{\beta\epsilon}^\alpha$ are the Christoffel symbols for the metric connection on $M$, are sufficient to guarantee this.

The metric connection on $M$ extends to a connection on $\bigwedge_\ast M$ via the Leibniz rule.  Moreover, the extended metric connection matrices $\hat{\Gamma}_\alpha$ satisfy both conditions in equation \eqref{eq:fullcompat}.  So to incorporate forces other than gravity, we look for connection matrices of the form
$$C_\alpha=\hat\Gamma_\alpha+\theta_\alpha,
  \quad\text{with}
$$
\begin{equation}\label{eq:physcon}
  \theta_\alpha^\dagger\hat{g}+\hat{g}\theta_\alpha=0
  \quad\text{and}\quad
  [\gamma^\alpha,\theta_\beta]=0.
\end{equation}
We call any collection $\theta_\alpha$ of $16\times 16$ matrices that satisfy \eqref{eq:physcon} a {\bf force tensor}.  We write $\boldsymbol\theta$ to denote the collection $\{\theta_0,\theta_1,\theta_2,\theta_3\}$.

By general principles, under a local change of basis $B$ for the tangent bundle of $M$, the connection matrices $C_\alpha$ transform as
\begin{equation*}
  C_\alpha'=(B^{-1})_\alpha^\beta(-\partial_\beta\hat{B}
                                  +\hat{B}C_\beta)\hat{B}^{-1}.
\end{equation*}
The extended metric connection $\hat\Gamma_\alpha$ satisfies this equation, so a force tensor must satisfy the transformation rule
\begin{equation}\label{eq:transform}
  \theta_\alpha'=(B^{-1})_\alpha^\beta\hat{B}\theta_\beta\hat{B}^{-1}.
\end{equation}

\section{The space of force tensors}\label{sec:space}

The set $V(g)$ of all force tensors is a real vector space.  We will determine a basis in the case of the metric $g=\eta$ on Minkowski space ${\amsmathbb M}$, where
$$\eta\doteq{\rm diag}(1,-1,-1,-1).$$

For a multi--index $I$, let $\Theta_I$ denote the $16\times 16$ matrix that commutes with gamma matrices and such that
$$\Theta_I{\bf e}_\emptyset={\bf e}_I.$$
We will see (Fact 1(c) below) that this completely determines $\Theta_I$.  It should be pointed out that even though $\Theta_\alpha{\bf e}_\emptyset={\bf e}_\alpha$ and $\gamma_\alpha{\bf e}_\emptyset={\bf e}_\alpha$, necessarily $\Theta_\alpha\neq\gamma_\alpha$.

The above theta matrices can be used to construct a basis of the space of force tensors $V(\eta)$.  Specifically, we let $\basis{\beta}{I}$ denote the collection of matrices with
$$(\basis{\beta}{I})_\alpha=0
  \,\,\text{if $\alpha\neq\beta$},
  \quad\text{and}\quad
  (\basis{\beta}{I})_\alpha=\Theta_I
  \,\,\text{if $\alpha=\beta$}.
$$
I.e., $(\basis{\beta}{I})_\alpha=\delta_\alpha^\beta\Theta_I$.  We will show that $V(\eta)$ has a basis given by
\begin{equation}\label{eq:basis}
  \begin{aligned}
    &\basis{\beta}{I}
       \quad\text{with}\quad
       I=0,1,2,3,01,02,03,12,13,23\\
    &i\basis{\beta}{J}
      \quad\text{with}\quad
      J=\emptyset,012,013,023,123,0123
  \end{aligned}
\end{equation}
and $\beta=0,1,2,3$.  In particular, $V(\eta)$ has dimension $64$.

\subsection{The space $V(\eta)$}

\begin{theorem}\label{fact:1}
Suppose $I=\alpha_1\cdots\alpha_k$ is a {\bf unique} multi--index.  That is, it contains no duplicate indices.
\begin{enumerate}[label=(\alph*)]
\item $\hat\eta$ is diagonal, and $\hat\eta_{II}=\eta_{\alpha_1\alpha_1}\cdots\eta_{\alpha_k\alpha_k}$.
\item Set $\gamma_I\doteq\gamma_{\alpha_1}\cdots\gamma_{\alpha_k}$ if $I\neq\emptyset$, and $\gamma_\emptyset\doteq{\mathcal I}$.  Then $\gamma_I{\bf e}_\emptyset={\bf e}_I$.
\item $\Theta_I{\bf e}_J=\gamma_J{\bf e}_I$ for all unique multi--indices $J$.
\item\label{lit:main} $\Theta_I^\dagger\hat\eta=s_I\hat\eta\Theta_I$, where $s_I=-1$ if $|I|=1,2$, and $s_I=1$ if $|I|=0,3,4$.
\end{enumerate}
\end{theorem}

\noindent
Here, ${\mathcal I}$ denotes the $16\times 16$ identity matrix.  It should be noted that if $I$ is not unique, then $\Theta_I=0$.  Moreover, the conclusion of (b) is false in this case.  The identity $\Theta_I{\bf e}_J=\gamma_J{\bf e}_I$ trivially holds if $I$ is not unique, but otherwise does not hold if $J$ is not unique.

\begin{theorem}
The space ${\mathbb C}\Theta$ of all $16\times 16$ matrices that commute with gamma matrices is a sixteen--dimensional complex vector space spanned by the matrices $\Theta_I$ for all unique multi--indices $I$.  Moreover, ${\mathbb C}\Theta$ is a matrix ring with unit $\Theta_\emptyset={\mathcal I}$, and forms a Clifford algebra representation: $\Theta_\alpha\Theta_\beta+\Theta_\beta\Theta_\alpha=-2\eta_{\alpha\beta}$.
\end{theorem}
\noindent

Item \ref{lit:main} of Fact \ref{fact:1} is equivalent to the statement: $\Theta_I^\dagger\hat\eta+\hat\eta\Theta_I=0$ for $|I|=1,2$, and $(i\Theta_J)^\dagger\hat\eta+\hat\eta(i\Theta_J)=0$ for $|J|=0,3,4$.  Thus the subspace of ${\mathbb C}\Theta$ consisting of matrices $N$ with $N^\dagger\hat\eta+\eta N=0$ is a sixteen--dimensional real vector space with basis $\Theta_I$ for $|I|=1,2$, and $i\Theta_J$ with $|J|=0,3,4$.  It follows that the space $V(\eta)$ of force tensors on Minkowski space indeed has a basis given by equation \eqref{eq:basis}.

\begin{proof}
All statements in both facts are straightforward, with the exception of Fact \ref{fact:1}\ref{lit:main}, for which we sketch an ad hoc argument.  First verify that $\gamma_I^\dagger\hat\eta=s_I\hat\eta\gamma_I$ using the identities $\gamma_\alpha^\dagger\hat\eta=-\hat\eta\gamma_\alpha$ and $\gamma_\alpha\gamma_\beta=-\gamma_\beta\gamma_\alpha$ if $\alpha\neq\beta$.  Set $N=z_I\Theta_I$, where $z_I=1$ for $|I|=1,2$ and $z_I=i$ for $|I|=0,3,4$.  Compute that ${\bf e}_K^\dagger(N^\dagger\hat\eta+\hat\eta N){\bf e}_\emptyset=\hat\eta_{KI}(z_I+s_K\bar{z}_I)=0$ for any unique $K$.  Now compute ${\bf e}_K^\dagger(N^\dagger\hat\eta+\hat\eta N){\bf e}_J=s_J{\bf e}_K^\dagger\gamma_J^\dagger(N^\dagger\hat\eta+\hat\eta N){\bf e}_\emptyset$ for any unique $J$, which must be zero by the previous computation.
\end{proof}

The vector space ${\mathbb C}\Theta$ has a secondary complex structure ${\mathcal J}$, given by the matrix
$${\mathcal J}\doteq\Theta_{0123}.$$
That is, ${\mathcal J}^2=-{\mathcal I}$.  The two complex structures combine to give a {\em perplex} structure $i{\mathcal J}$: $(i{\mathcal J})^2={\mathcal I}$.  Observe that ${\mathcal J}$ induces a complex structure on two subspaces of $V(\eta)$.  The first is the subspace spanned by force tensors of the form $i\basis{\alpha}{I}$ where $|I|=0,4$.  And the second is the subspace spanned by $\basis{\alpha}{I}$ with $|I|=2$.  On the other hand, the subspace spanned by $\basis{\alpha}{I}$ and $i\basis{\beta}{J}$ with $|I|=1$ and $|J|=3$ has a perplex structure induced by $i{\mathcal J}$.

\subsection{Theta matrix identities}

The gamma and theta matrices both satisfy the Clifford algebra relation, so the usual gamma matrix algebra and trace identities will also be satisfied by the theta matrices.  In fact, the gamma and theta matrices are similar in the sense that $\Theta_\alpha=S\gamma_\alpha S^{-1}$.  Indeed, $S$ can be taken to be diagonal with $S_I^I=1$ if $|I|=0,1,4$, and $S_I^I=-1$ if $|I|=2,3$.

We state some useful identities specific to theta matrices.  Here, $\epsilon_{\alpha\beta\rho\sigma}$ is the totally anti--symmetric Levi--Civita symbol.
\begin{align*}
  \Theta_I{\mathcal J}=(-1)^{|I|}{\mathcal J}\theta_I
  & &
  {\mathcal J}\Theta_{\alpha\beta\rho}
    =-{\epsilon_{\alpha\beta\rho}}^\sigma\Theta_\sigma
  & &
  {\mathcal J}\Theta_{\alpha\beta}
     =\tfrac{1}{2}{\epsilon_{\alpha\beta}}^{\rho\sigma}\Theta_{\rho\sigma}
\end{align*}
\begin{align*}
  \Theta_{\alpha\beta\rho\sigma}=\epsilon_{\alpha\beta\rho\sigma}{\mathcal J}
  & &
  {\it tr}(\Theta_I)=16\,\delta_{I\emptyset}
  & &
  {\it tr}(\Theta_\alpha\Theta_{\rho\sigma})=0
  & &
  {\it tr}({\mathcal J}\Theta_\alpha\Theta_{\rho\sigma})=0
\end{align*}
\begin{align*}
  {\it tr}(\Theta_{\alpha\beta}\Theta_{\rho\sigma})
    =16(\eta_{\alpha\sigma}\eta_{\beta\rho}-\eta_{\alpha\rho}\eta_{\beta\sigma})
  & &
  {\it tr}({\mathcal J}\Theta_{\alpha\beta}\Theta_{\rho\sigma})
     =-16\,\epsilon_{\alpha\beta\rho\sigma}
\end{align*}
\begin{align*}
  \Theta_\alpha\Theta^\alpha=-4\,{\mathcal I}
  & &
  \Theta_\alpha\Theta_\beta=-\Theta_{\alpha\beta}-\eta_{\alpha\beta}{\mathcal I}
\end{align*}
\begin{align*}
  \Theta^\alpha\Theta_{\alpha\beta}=\Theta_{\beta\alpha}\Theta^\alpha=3\Theta_\beta
  & &
  \eta^{\rho\sigma}\Theta_{\alpha\rho}\Theta_{\sigma\beta}
     =2\Theta_{\alpha\beta}+3\eta_{\alpha\beta}{\mathcal I}
\end{align*}
\begin{align*}
  \eta^{\rho\sigma}\Theta_{\alpha\rho}\Theta_{\mu\nu}\Theta_{\sigma\beta}
   =-\Theta_{\alpha\beta}\Theta_{\mu\nu}-\Theta_{\mu\nu}\Theta_{\alpha\beta}
    -\eta_{\alpha\beta}\Theta_{\mu\nu}
\end{align*}
\begin{align*}
  [\Theta_{\alpha\beta},\Theta_{\rho\sigma}]
     =2\eta_{\beta\rho}\Theta_{\alpha\sigma}
     +2\eta_{\alpha\sigma}\Theta_{\beta\rho}
     -2\eta_{\alpha\rho}\Theta_{\beta\sigma}
     -2\eta_{\beta\sigma}\Theta_{\alpha\rho}
\end{align*}

\section{Irreducible force tensors}\label{sec:irred}

A Lorentz transformation $\Lambda$ defines a change of basis for the tangent bundle of $M$.  We can extend this to a change of basis $\hat\Lambda$ for $\bigwedge_\ast M$.  In this way the Lorentz group $O(g)$ acts on the space of force tensors $V(g)$.  Indeed, from equation \eqref{eq:transform}, 
$$(\Lambda\cdot\boldsymbol\theta)_\alpha
  \doteq(\Lambda^{-1})_\alpha^\beta\hat{\Lambda}\theta_\beta\hat{\Lambda}^{-1}
$$
for any force tensor $\boldsymbol\theta$.  The corresponding Lorentz algebra action is then
\begin{equation}\label{eq:action}
  (L\cdot\boldsymbol\theta)_\alpha
  =-L_\alpha^\beta\theta_\beta
   +\hat{L}\theta_\alpha -\theta_\alpha\hat{L}
  =[\hat{L},\theta_\alpha]-L_\alpha^\beta \theta_\beta
\end{equation}
for $L\in so(g)$.

We say that a force tensor $\boldsymbol\theta$ is {\bf irreducible} if its orbit under the $O(g)$ action, or equivalently under the $so(g)$ action, lies in an irreducible real representation of $O(g)$.  The irreducible sub--representations of $V(g)$ can be computed using standard Lie algebra techniques, such as in \cite{Fulton}.  In this section, we will present the results for the case $g=\eta$.

Let us first indicate explicitly how ${\it so}(\eta)$ affects each of the basic force tensors in equation \eqref{eq:basis}.  The Lorentz algebra ${\it so}(\eta)$ is a real six--dimensional vector space consisting of $4\times 4$ matrices $L$ that satisfy $L^T\eta+\eta L=0$.  From equation \eqref{eq:action}, the action of $L\in so(\eta)$ on the force tensor $\basis{\beta}{I}$ is given by
\begin{equation}\label{eq:basisaction}
  (L\cdot\basis{\beta}{I})_\alpha
  =[\hat{L},(\basis{\beta}{I})_\alpha]
   -L_\alpha^\nu(\basis{\beta}{I})_\nu
  =\delta_\alpha^\beta[\hat{L},\Theta_I]
   -L_\alpha^\beta\Theta_I.
\end{equation}
In \cite{me} it is shown that $\hat\Lambda\gamma_\alpha\hat\Lambda^{-1}=\Lambda_\alpha^\beta\gamma_\beta$ for any $\Lambda\in O(\eta)$.  It follows that $[\hat{L},\gamma_\alpha]=L_\alpha^\beta\gamma_\beta$.  And with this, one shows that $[\hat{L},\Theta_I]$ commutes with gamma matrices.  So that its value is determined by its effect on ${\bf e}_\emptyset$, which can be computed using equation \eqref{eq:liealg}.

As an example, we compute $L\cdot\basis{\beta}{\beta\rho}$.  First,
$$[\hat{L},\Theta_{\beta\rho}]{\bf e}_\emptyset
  =\hat{L}\Theta_{\beta\rho}{\bf e}_\emptyset
  =\hat{L}{\bf e}_{\beta\rho}
  =L_\beta^\mu{\bf e}_{\mu\rho}+L_\rho^\mu{\bf e}_{\beta\mu}
  =(L_\beta^\mu\Theta_{\mu\rho}
    +L_\rho^\mu\Theta_{\beta\mu}){\bf e}_\emptyset
$$
(recall that $\hat{L}{\bf e}_\emptyset=0$).  Whence $[\hat{L},\Theta_{\beta\rho}]=L_\beta^\mu\Theta_{\mu\rho}+L_\rho^\mu\Theta_{\beta\mu}$.  By equation \eqref{eq:basisaction}, we then have
$$(L\cdot\basis{\beta}{\beta\rho})_\alpha
  =L_\alpha^\mu\Theta_{\mu\rho}+L_\rho^\mu\Theta_{\alpha\mu}
   -L_\alpha^\beta\Theta_{\beta\rho}
  =L_\rho^\mu\Theta_{\alpha\mu}
  =(L_\rho^\mu\basis{\beta}{\beta\mu})_\alpha.
$$
That is, $L\cdot\basis{\beta}{\beta\rho}=L_\rho^\mu\basis{\beta}{\beta\mu}$, so that $\basis{\beta}{\beta\rho}$ transforms like ${\bf e}_\rho$.  In general, one shows that $\basis{\beta}{\beta\rho_1\cdots\rho_k}$ transforms like ${\bf e}_{\rho_1\cdots\rho_k}$.

\subsection{Irreducible summands: overview}

Viewing the space of force tensors $V(g)$ as a Lie algebra representation of $so(g)$, we decompose it into irreducible summands.  Schematically, the decomposition is
\begin{equation}\label{eq:summands}
  \begin{aligned}
    V(g)&=V_r(g)\oplus iV_c(g)\\
    V_r(g)&=\mathbb{1}\oplus\mathbb{4}\oplus\mathbb{4}'
            \oplus\mathbb{6}\oplus\mathbb{9}\oplus{\mathbb{1}\mathbb{6}}\\
    V_c(g)&=\mathbb{1}'\oplus\mathbb{4}''\oplus\mathbb{4}'''
            \oplus\mathbb{6}'\oplus\mathbb{9}'
  \end{aligned}
\end{equation}
where each summand has the indicated dimension.  All summands of the same dimension are isomorphic.  The dimension four summands are isomorphic to $(\tfrac{1}{2},\tfrac{1}{2})$ in the classification of representations of the Lorentz group.  The dimension six, nine, and sixteen summands are isomorphic to $(1,0)\oplus(0,1)$, $(1,1)$, and $(\tfrac{3}{2},\tfrac{1}{2})\oplus(\tfrac{1}{2},\tfrac{3}{2})$, respectively.  While the summands $\mathbb{6}$ and $\mathbb{1}\mathbb{6}$ are reducible over the complex numbers, they are irreducible over the reals.

In the case when $g=\eta$, we can be more specific about the summands in \eqref{eq:summands}.  One computes that the complex structure matrix ${\mathcal J}$ commutes with the extended Lorentz Lie algebra action.  So that if we define ${\mathcal J}$ to act component--wise on $\basis{\alpha}{I}$, that is $({\mathcal J}\basis{\alpha}{I})_\beta=\delta_\beta^\alpha{\mathcal J}\Theta_I$, then $L\cdot({\mathcal J}\basis{\alpha}{I})={\mathcal J}(L\cdot\basis{\alpha}{I})$.  And we have
$$\mathbb{1}'={\mathcal J}\mathbb{1},\quad
  \mathbb{6}'={\mathcal J}\mathbb{6},\quad
  \mathbb{4}'={\mathcal J}\mathbb{4},\quad
  \mathbb{4}'''={\mathcal J}\mathbb{4}'',\quad
  \mathbb{9}'={\mathcal J}\mathbb{9}$$
as representations of ${\it so}(\eta)$.  Whence
\begin{equation}\label{eq:summands-alt}
\begin{aligned}
  V(\eta)
  &=(1+i{\mathcal J})\mathbb{1}
    \oplus(1+{\mathcal J})\mathbb{4}
    \oplus i(1+{\mathcal J})\mathbb{4}''\\
  &\quad\quad
    \oplus(1+i{\mathcal J})\mathbb{6}
    \oplus(1+i{\mathcal J})\mathbb{9}
    \oplus\mathbb{1}\mathbb{6}.
\end{aligned}
\end{equation}
Each of the summands will be described explicitly in the remainder of this section.

\subsection{One--dimensional force tensors}\label{sec:con1}

The summand $\mathbb{1}$ is the one--dimensional subspace spanned by the single force tensor
$${\bf u}\doteq\basis{\alpha}{\alpha}.
$$
Indeed, $L\cdot{\bf u}=0$ for all $L\in so(\eta)$.  Any force tensor in $(1+i{\mathcal J})\mathbb{1}$ can thus be written in the form ${\bf U}=\phi(\cos\zeta+i\sin\zeta\,{\mathcal J}){\bf u}$ for a real scalar field $\phi$, and constant angle $\zeta$.  The individual component matrices of ${\bf U}$ are
\begin{equation}\label{eq:1-connection}
  U_\alpha=\phi(\cos\zeta+i\sin\zeta\,{\mathcal J})\Theta_\alpha.
\end{equation}

{\bf Remark.}
For all compound summands in $V(\eta)$, such as $(1+i{\mathcal J})\mathbb{1}$, there will be an angular parameter $\zeta$.  It gives a free parameter of the model, and although we use the same symbol, its value is not the same across distinct compound summands.

\subsection{Real four--dimensional force tensors}\label{ssec:4dcon}

The four--dimensional summand $\mathbb{4}$ in \eqref{eq:summands} has a basis given by the force tensors
$${\bf v}_\alpha\doteq\basis{\beta}{\beta\alpha}.$$
As we have seen, these force tensors transform exactly like the basis vectors ${\bf e}_\alpha$ of ${\amsmathbb M}$ under Lorentz transformations.  Any force tensor in $(1+{\mathcal J})\mathbb{4}$ can therefore be written in the form ${\bf F}=B^\alpha(\cos\zeta+\sin\zeta\,{\mathcal J}){\bf v}_\alpha=B^\alpha e^{\zeta {\mathcal J}}\,{\bf v}_\alpha$ for some 4--vector field $B^\alpha$ and fixed angle $\zeta$.  Here $e^{\zeta{\mathcal J}}$ is the matrix exponential
$$e^{\zeta{\mathcal J}}=\cos\zeta\,{\mathcal I}+\sin\zeta\,{\mathcal J}.$$
The components of ${\bf F}$ can be written as
\begin{equation}\label{eq:4-connection}
  F_\alpha=B^\beta e^{\zeta{\mathcal J}}\Theta_{\alpha\beta}.
\end{equation}

\subsection{Imaginary four--dimensional force tensors}\label{ssec:i4dcon}

The four--dimensional summand $i\mathbb{4}''$ in \eqref{eq:summands} has a basis consisting of the force tensors
$${\bf w}_\alpha\doteq i\eta_{\alpha\beta}\basis{\beta}{\emptyset}.$$
One computes that ${\it so}(\eta)$ also acts on the force tensors ${\bf w}_\alpha$ in exactly the same way as on the vectors ${\bf e}_\alpha$.  Consequently, any force tensor in $i(1+{\mathcal J})\mathbb{4}''$ is of the form ${\bf M}=A^\beta e^{\zeta{\mathcal J}}{\bf w}_\beta$ for some 4--vector field $A^\alpha$. In other words, the component matrices of ${\bf M}$ are
\begin{equation}\label{eq:4-connection2}
  M_\alpha=iA_\alpha e^{\zeta{\mathcal J}}.
\end{equation}

\subsection{Six--dimensional force tensors}

The summand $\mathbb{6}$ in \eqref{eq:summands} is spanned by the force tensors
$${\bf v}_{\alpha\beta}={\mathcal J}{\bf w}_{\alpha\beta},
  \quad\text{for $\alpha<\beta$,}
  \quad\text{where}\quad
  {\bf w}_{\alpha\beta}\doteq\basis{\epsilon}{\epsilon\alpha\beta}.
$$
The action of ${\it so}(\eta)$ on ${\bf w}_{\alpha\beta}$, and consequently on ${\bf v}_{\alpha\beta}$ as well, is the same as on ${\bf e}_{\alpha\beta}$.  Hence any force tensor from $(1+i{\mathcal J})\mathbb{6}$ can be written as ${\bf H}=W^{\alpha\beta}(\cos\zeta +i\sin\zeta\,{\mathcal J}){\bf v}_{\alpha\beta}$, for some anti--symmetric tensor field $W^{\alpha\beta}$.  That is, the component matrices of ${\bf H}$ are
\begin{equation}\label{eq:6-connection}
  H_\alpha=-iW^{\rho\sigma}
              (\sin\zeta+i\cos\zeta\,{\mathcal J})
              \Theta_{\alpha\rho\sigma}.
\end{equation}

\subsection{Nine--dimensional force tensors}

Let $\mathbb{1}\mathbb{0}$ be the ten--dimensional real vector space spanned by the force tensors
$${\bf u}_{\alpha\beta}
  \doteq\tfrac{1}{2}\bigl(\eta_{\alpha\epsilon}\basis{\epsilon}{\beta}
                          +\eta_{\beta\epsilon}\basis{\epsilon}{\alpha}\bigr)
  \quad\text{for $\alpha\leq\beta$}.
$$
Necessarily, $\mathbb{1}\mathbb{0}$ is a subspace of $V(\eta)$.  We have ${\bf u}_{\alpha\beta}={\bf u}_{\beta\alpha}$, and in fact we may identify $\mathbb{1}\mathbb{0}$ with the second symmetric power $S^2(\amsmathbb{M})$ of $\amsmathbb{M}$.  Indeed, one computes that the ${\it so}(\eta)$ action is given by $L\cdot{\bf u}_{\alpha\beta}=L_\alpha^\epsilon{\bf u}_{\epsilon\beta}+L_\beta^\epsilon{\bf u}_{\alpha\epsilon}$ for any $L\in{\it so}(\eta)$.  Therefore, a force tensor on $\mathbb{1}\mathbb{0}$ is of the form $S^{\alpha\beta}{\bf u}_{\alpha\beta}$ for some symmetric tensor field $S^{\alpha\beta}$ on $\amsmathbb{M}$.

Observe that $\eta^{\alpha\beta}{\bf u}_{\alpha\beta}=\basis{\alpha}{\alpha}={\bf u}$, so that $\mathbb{1}$ is a subspace of $\mathbb{1}\mathbb{0}$.  Moreover, the map $\pi:\mathbb{1}\mathbb{0}\rightarrow\mathbb{1}$ given by $\pi(S^{\alpha\beta}{\bf u}_{\alpha\beta})=\tfrac{1}{4}S_\alpha^\alpha{\bf u}$, where $S_\alpha^\alpha\doteq\eta_{\alpha\beta}S^{\alpha\beta}$, is a projection.  The space $\mathbb{9}$ is the complement of $\mathbb{1}$ in $\mathbb{1}\mathbb{0}$.  That is, a force tensor on $\mathbb{9}$ is of the form $S^{\alpha\beta}{\bf u}_{\alpha\beta}$, where $S^{\alpha\beta}$ is now symmetric and contraction--free.  I.e., $S_\alpha^\alpha=0$.

From the above argument, a force tensor in $(1+i{\mathcal J})\mathbb{9}$ can be written as ${\bf N}=S^{\alpha\beta}(\cos\zeta+i\sin\zeta\,{\mathcal J}){\bf u}_{\alpha\beta}$, which has component matrices
\begin{equation}\label{eq:9-connection}
  N_\alpha=\eta_{\alpha\beta} S^{\beta\epsilon}
            (\cos\zeta+i\sin\zeta\,{\mathcal J})\Theta_\epsilon
\end{equation}
for $S^{\alpha\beta}$ a symmetric and contraction--free real field on $\amsmathbb{M}$.

\subsection{Sixteen--dimensional force tensors}

Consider the 24--dimensional subspace $\mathbb{2}\mathbb{4}$ of $V(\eta)$ spanned by the force tensors
$${\bf u}_{\alpha\rho\sigma}
  \doteq\eta_{\alpha\beta}\basis{\beta}{\rho\sigma}
  \quad\text{with $\rho<\sigma$.}
$$
One verifies that $L\cdot{\bf u}_{\alpha\rho\sigma}=L_\alpha^\epsilon{\bf u}_{\epsilon\rho\sigma}+L_\rho^\epsilon{\bf u}_{\alpha\epsilon\sigma}+L_\sigma^\epsilon{\bf u}_{\alpha\rho\epsilon}$ for any $L\in{\it so}(\eta)$.  Consequently as ${\bf u}_{\alpha\rho\sigma}=-{\bf u}_{\alpha\sigma\rho}$, $\mathbb{2}\mathbb{4}$ can be identified with the tensor product $\amsmathbb{M}\otimes\bigwedge^2(\amsmathbb{M})$ of Minkowski space and its second exterior power.  Thus any force tensor in $\mathbb{2}\mathbb{4}$ can be written as $F^{\alpha\rho\sigma}{\bf u}_{\alpha\rho\sigma}$ for some tensor field $F^{\alpha\rho\sigma}$ on $\amsmathbb{M}$ with $F^{\alpha\rho\sigma}=-F^{\alpha\sigma\rho}$.

Because $\eta^{\alpha\rho}{\bf u}_{\alpha\rho\sigma}=\basis{\beta}{\beta\sigma}={\bf v}_\sigma$, we see that $\mathbb{4}$ is a subspace of $\mathbb{2}\mathbb{4}$.  And consequently, so is ${\mathcal J}\mathbb{4}$.  In fact, we have that $\epsilon^{\alpha\rho\sigma\tau}{\bf u}_{\alpha\rho\sigma}=2\eta^{\tau\beta}{\mathcal J}{\bf v}_\beta$.  The projections $\pi_1:\mathbb{2}\mathbb{4}\rightarrow\mathbb{4}$ and $\pi_2:\mathbb{2}\mathbb{4}\rightarrow{\mathcal J}\mathbb{4}$ are given by
$$\pi_1(F^{\alpha\rho\sigma}{\bf u}_{\alpha\rho\sigma})
   =\tfrac{1}{4}\eta_{\alpha\rho}F^{\alpha\rho\sigma}{\bf v}_\sigma
   \quad\text{and}\quad
   \pi_2(F^{\alpha\rho\sigma}{\bf u}_{\alpha\rho\sigma})=\tfrac{1}{3}\eta^{\mu\nu}\epsilon_{\nu\alpha\rho\sigma}F^{\alpha\rho\sigma}{\mathcal J}{\bf v}_\mu.
$$
One computes that $\pi_1\circ\pi_2=0$, so that the subspaces $\mathbb{4}$ and ${\mathcal J}\mathbb 4$ are orthogonal.   Thus $(1{+\mathcal J})\mathbb{4}$ forms an 8--dimensional subspace of $\mathbb{2}\mathbb{4}$.

The space $\mathbb{1}\mathbb{6}$ is the complement of $(1+{\mathcal J})\mathbb{4}$ in $\mathbb{2}\mathbb{4}$.  It follows that any force tensor in $\mathbb{1}\mathbb{6}$ is of the form ${\bf X}=F^{\alpha\rho\sigma}{\bf u}_{\alpha\rho\sigma}$ where $F^{\alpha\rho\sigma}$ satisfies
\begin{equation}\label{eq:fierz}
  F^{\alpha\rho\sigma}=-F^{\alpha\sigma\rho},
  \quad
  \eta_{\alpha\rho}F^{\alpha\rho\sigma}=0,
  \quad
  \epsilon_{\alpha\rho\sigma\tau}F^{\alpha\rho\sigma}=0.
\end{equation}
Tensors $F^{\alpha\rho\sigma}$ that satisfy \eqref{eq:fierz} are called (self--dual) {\em Fierz tensors,} and are used in spin--2 field theory, see \cite{Novello}.  The component matrices of ${\bf X}$ are
\begin{equation}\label{eq:16-connection}
  X_\alpha=\eta_{\alpha\beta}F^{\beta\rho\sigma}\Theta_{\rho\sigma}.
\end{equation}

\section{Fundamental forces}\label{sec:force}

Each distinct irreducible summand in \eqref{eq:summands-alt} corresponds to a distinct fundamental force.  That is, there are six fundamental forces (not including gravity) allowed by this model.  In this section, for each fundamental force we find the potential energy under the assumption
$$L_V\doteq\tfrac{1}{2}\omega\,{\it tr}(\Omega_{\alpha\beta}\Omega^{\alpha\beta})
     +c.c.,$$
where $\Omega_{\alpha\beta}=\partial_\alpha C_\beta-\partial_\beta C_\alpha+C_\alpha C_\beta-C_\beta C_\alpha$ is the curvature matrix of the total connection $C_\alpha=\hat\Gamma_\alpha+\theta_\alpha$, with $\boldsymbol\theta$ a force tensor.  We also give the field equations obtained by varying the Lagrangian density
\begin{equation}\label{eq:lagrangian}
  L=L_K-\tau L_V
\end{equation}
with respect to the field components.  Here $L_K$ is as in equation \eqref{eq:diraclagrangian}, and $\tau$ is a constant.  Again we assume that $g=\eta$, so that $C_\alpha=\theta_\alpha$.

\subsection{Scalar field}

The potential energy of the force tensor ${\bf U}$ in equation \eqref{eq:1-connection}, associated with the summand $(1+i{\mathcal J})\mathbb{1}$, is found to be
\begin{equation}\label{eq:phi4}
  L_V=-96\cos{2\zeta}\,[(\partial^\alpha\phi)(\partial_\alpha\phi)
                        +8\cos{2\zeta}\,\phi^4].
\end{equation}
Variation of the Lagrangian density \eqref{eq:lagrangian} with respect to the real scalar field $\phi$ then gives the field equation
\begin{equation}
  \partial_\alpha\partial^\alpha\phi-16\cos{2\zeta}\,\phi^3
  =\tfrac{1}{192\tau\cos{2\zeta}}\psi^\dagger\hat\eta\gamma^\alpha
      (\cos\zeta+i\sin\zeta\,{\mathcal J})\Theta_\alpha\psi.
\end{equation}
Like $\zeta$, $\tau$ is a free parameter of the model, and is not constant across the distinct compound summands in \eqref{eq:summands-alt}.

\subsection{Vector fields}

There are two compound summands of dimension four in \eqref{eq:summands-alt}.  Each will give rise to a different 4--vector field.

\subsubsection{Vector field of the first type}

The potential energy of the force tensor ${\bf F}$ associated to the summand $(1+{\mathcal J})\mathbb{4}$ given in equation \eqref{eq:4-connection} is
\begin{equation}
\begin{aligned}
  L_V=&-384(B_\alpha B^\alpha)^2\cos{4\zeta}
        -256(B^\alpha B_\alpha\partial_\beta B^\beta
             -B^\alpha B_\beta\partial_\alpha B^\beta)\cos{3\zeta}\\
     &-32[2(\partial_\alpha B^\beta)(\partial^\alpha B_\beta)
               +(\partial_\alpha B^\alpha)^2]\cos{2\zeta}\\
     &-32\epsilon_{\alpha\beta\rho\sigma}
          (\partial^\alpha B^\beta)(\partial^\rho B^\sigma)\sin{2\zeta}.
\end{aligned}
\end{equation}
The last summand is a divergence, and has no effect on the field variation.  The field equation for the real four--vector field $B^\alpha$ is then
\begin{equation}\label{eq:4field}
\begin{aligned}
  &\partial_\beta\partial^\beta B_\alpha
   +\tfrac{1}{2}\partial_\alpha\partial_\beta B^\beta
   +\tfrac{6\cos{3\zeta}}{\cos{2\zeta}}
      (B_\beta\partial_\alpha B^\beta-B_\alpha\partial_\beta B^\beta)\\
  &\quad\quad
   -\tfrac{12\cos{4\zeta}}{\cos{2\zeta}}(B_\beta B^\beta)B_\alpha
   =\tfrac{1}{128\tau\cos{2\zeta}}\psi^\dagger\hat\eta\gamma^\beta
        e^{\zeta{\mathcal J}}\Theta_{\beta\alpha}\psi.
\end{aligned}
\end{equation}

\subsubsection{Vector field of the second type}

On the other hand, for the force tensor ${\bf M}$ on $i(1+\mathcal{J})\mathbb{4}''$ in equation \eqref{eq:4-connection2}, we have the potential energy
\begin{equation}
  L_V=-32\cos{2\zeta}\,[(\partial_\alpha A^\beta)(\partial^\alpha A_\beta)
                          -(\partial_\alpha A^\beta)(\partial_\beta A^\alpha)]
\end{equation}
for the real 4--vector field $A^\alpha$.  This leads to the field equation
\begin{equation}\label{eq:4fieldi}
  \partial_\beta\partial^\beta A_\alpha
  -\partial_\alpha\partial_\beta A^\beta
  =\tfrac{i}{64\tau\cos{2\zeta}}\psi^\dagger\hat\eta
     \gamma_\alpha e^{\zeta{\mathcal J}}\psi.
\end{equation}

\subsection{Antisymmetric tensor field}

In this case, the potential energy term of the Lagrangian for the force tensor ${\bf H}$ on $(1+i{\mathcal J})\mathbb{6}$ in equation \eqref{eq:6-connection} is
\begin{equation}
\begin{aligned}
  L_V
  &=1024\cos^2{2\zeta}\,[W^{\alpha\beta}W_{\beta\rho}W^{\rho\sigma}W_{\sigma\alpha}
                         -(W^{\alpha\beta}W_{\alpha\beta})^2]\\
  &\quad\quad
      \mbox{}+64\cos{2\zeta}\,[(\partial_\alpha W^{\rho\sigma})
                               (\partial^\alpha W_{\rho\sigma})
                              +2(\partial_\beta W^{\alpha\beta})
                                (\partial^\sigma W_{\alpha\sigma})].
\end{aligned}
\end{equation}
The field equation for the real antisymmetric tensor field $W^{\alpha\beta}$ is then
\begin{equation}
\begin{aligned}
  &\partial_\mu\partial^\mu W_{\alpha\beta}
   -\partial_\alpha\partial^\sigma W_{\beta\sigma}
   +\partial_\beta\partial^\sigma W_{\alpha\sigma}\\
  &\quad\quad
   \mbox{}
   +16\cos{2\zeta}\,
      [W_{\alpha\rho}W^{\rho\sigma}W_{\sigma\beta}
       +(W^{\rho\sigma}W_{\rho\sigma})W_{\alpha\beta}]\\
  &=\tfrac{i}{128\tau\cos{2\zeta}}\psi^\dagger\hat\eta
      \gamma^\nu(\sin\zeta+i\cos\zeta\,{\mathcal J})
      \Theta_{\nu\alpha\beta}\psi.
\end{aligned}
\end{equation}

\subsection{Symmetric contraction--free tensor field}

The potential energy of the force tensor ${\bf N}$ on the summand $(1+i{\mathcal J})\mathbb{9}$, equation \eqref{eq:9-connection}, is given by
\begin{equation}
\begin{aligned}
  L_V&=64\cos^2{2\zeta}\,[S^{\alpha\beta}S_{\beta\rho}S^{\rho\sigma}S_{\sigma\alpha}
                          -(S^{\alpha\beta}S_{\alpha\beta})^2]\\
     &\quad\quad
      -32\cos{2\zeta}\,[ (\partial_\alpha S^{\beta\rho})
                          (\partial^\alpha S_{\beta\rho})
                        -(\partial^\alpha S^{\beta\rho})
                          (\partial_\beta S_{\alpha\rho})].
\end{aligned}
\end{equation}
To compute the variation of the total Lagrangian with respect to $S^{\alpha\beta}$, we need to take into account the constraint $S_\rho^\rho=\eta_{\rho\sigma}S^{\rho\sigma}=0$.  The result is
\begin{equation}
\begin{aligned}
  &\partial_\mu\partial^\mu S_{\alpha\beta}
   -\tfrac{1}{2}\partial_\alpha\partial^\mu S_{\mu\beta}
   -\tfrac{1}{2}\partial_\beta\partial^\mu S_{\mu\alpha}
   +\tfrac{1}{4}\eta_{\alpha\beta}\partial_\mu\partial_\nu S^{\mu\nu}\\
  &\quad\quad
   +4\cos{2\zeta}\,[S_{\alpha\mu}S^{\mu\nu}S_{\nu\beta}
                    -(S^{\mu\nu}S_{\mu\nu})S_{\alpha\beta}
                    -\tfrac{1}{4}\eta_{\alpha\beta}S_\mu^\nu S_\nu^\sigma S_\sigma^\mu]\\
  &=\tfrac{1}{128\tau\cos{2\zeta}}
    \psi^\dagger\hat\eta(\cos\zeta+i\sin\zeta\,{\mathcal J})
                (\gamma_\alpha\Theta_\beta
                 +\gamma_\beta\Theta_\alpha
                 -\tfrac{1}{2}\eta_{\alpha\beta}\gamma^\mu\Theta_\mu)\psi
\end{aligned}
\end{equation}
with $S^{\alpha\beta}$ a real symmetric contraction--free tensor field.

\subsection{Fierz tensor field}

Here the potential energy of the term of the force tensor ${\bf X}$ in equation \eqref{eq:16-connection} is given by
\begin{equation}
  L_V=-32F_{\alpha\beta\rho\sigma}F^{\alpha\beta\rho\sigma},
\end{equation}
where
$$F_{\alpha\beta\rho\sigma}
  \doteq \partial_\alpha F_{\beta\rho\sigma}
        -\partial_\beta F_{\alpha\rho\sigma}
        +4( {F_{\alpha\rho}}^\nu F_{\beta\nu\sigma}
           -{F_{\beta\rho}}^\nu F_{\alpha\nu\sigma})
$$
and $F^{\alpha\rho\sigma}$ is a Fierz tensor: a tensor that satisfies equation \eqref{eq:fierz}.  Constrained variation of the total Lagrangian yields
\begin{equation}
\begin{aligned}
  &\xi_{\alpha\rho\sigma}
  -\tfrac{1}{3}\eta_{\alpha\rho}{\xi^\beta}_{\beta\sigma}
  +\tfrac{1}{3}\eta_{\alpha\sigma}{\xi^\beta}_{\beta\rho}
  -\tfrac{1}{6}\epsilon_{\alpha\rho\sigma\omega}\epsilon^{\lambda\mu\nu\omega}
               \xi_{\lambda\mu\nu}\\
  &\quad\quad\quad\quad
   =\chi_{\alpha\rho\sigma}
   -\tfrac{1}{3}\eta_{\alpha\rho}{\chi^\beta}_{\beta\sigma}
   +\tfrac{1}{3}\eta_{\alpha\sigma}{\chi^\beta}_{\beta\rho}
   -\tfrac{1}{6}\epsilon_{\alpha\rho\sigma\omega}\epsilon^{\lambda\mu\nu\omega}
                \chi_{\lambda\mu\nu}
\end{aligned}
\end{equation}
where $\xi_{\alpha\rho\sigma}=\tfrac{1}{2}(\tilde\xi_{\alpha\rho\sigma}-\tilde\xi_{\alpha\sigma\rho})$ and
\begin{align*}
  \tilde\xi_{\alpha\rho\sigma}
   &\doteq 8(2{F^{\beta\nu}}_\rho\,\partial_\beta F_{\alpha\sigma\nu}
             +{F^{\beta\nu}}_\rho\,\partial_\alpha F_{\beta\nu\sigma}
             +{F_{\alpha\rho}}^\nu\partial_\beta{F^\beta}_{\sigma\nu})\\
   &\quad\quad\mbox{}
         -32(F_{\beta\rho\mu}{F_\alpha}^{\mu\nu}{F^\beta}_{\nu\sigma}
             -F_{\beta\rho\mu}F^{\beta\mu\nu}F_{\alpha\nu\sigma})
         +\partial_\beta\partial^\beta F_{\alpha\rho\sigma}
          -\partial_\beta\partial_\alpha{F^\beta}_{\rho\sigma}\\
  \chi_{\alpha\rho\sigma}
   &\doteq\tfrac{1}{128\tau}
          \psi^\dagger\hat\eta\gamma_\alpha\Theta_{\rho\sigma}\psi.
\end{align*}




\end{document}